# A Prediction Study of Path Loss Models from 2-73.5 GHz in an Urban-Macro Environment

Timothy A. Thomas[a], Marcin Rybakowski[b], Shu Sun[c], Theodore S. Rappaport[c], Huan Nguyen[d], István Z. Kovács[e], Ignacio Rodriguez[d]

[a]Nokia, Arlington Heights, IL, 60004, USA, [b]Nokia, Wroclaw, Poland

[c]NYU WIRELESS and Tandon School of Engineering, New York University, Brooklyn, NY 11201, USA

[d]Aalborg University, Aalborg, Denmark 9220, [e]Nokia, Aalborg Denmark 9220

Corresponding author: timothy.thomas@nokia.com

***Abstract*—It is becoming clear that 5G wireless systems will encompass frequencies from around 500 MHz all the way to around 100 GHz. To adequately assess the performance of 5G systems in these different bands, path loss (PL) models will need to be developed across this wide frequency range. The PL models can roughly be broken into two categories, ones that have some anchor in physics, and ones that curve-match only over the data set without any physical anchor. In this paper we use both real-world measurements from 2 to 28 GHz and ray-tracing studies from 2 to 73.5 GHz, both in an urban-macro environment, to assess the prediction performance of the two PL modeling techniques. In other words, we look at how the two different PL modeling techniques perform when the PL model is applied to a prediction set which is different in distance, frequency, or environment from a measurement set where the parameters of the respective models are determined. We show that a PL model with a physical anchor point can be a better predictor of PL performance in the prediction sets while also providing a parameterization which is more stable over a substantial number of different measurement sets.**

## I. INTRODUCTION

Recently there has been great interest in using the higher frequencies (e.g., beyond 6 GHz) for access communications [1]-[4]. The interest in these bands is growing because of the rapidly increasing use of wireless data by consumers coupled with the limited spectrum available at the lower frequencies (e.g., below 6 GHz). However before these higher frequency bands can be utilized, accurate path loss (PL) models must be determined for these bands like was done at lower frequencies (e.g., [5]) and preliminary work is under way [6]-[9].

These PL models will need to be developed for multiple environments including urban-macro (UMa). Note that even at the higher frequencies (e.g., >30 GHz), cell coverage (or at least partial coverage with adequate reflections) could still be possible especially for UMa scenarios similar to the 3GPP criteria of a 500 m inter-site distance (i.e., only requiring a 250 m cell range) [10]. This coverage could be possible at the higher frequency bands through very large antenna arrays (e.g., 100 elements or more) which are feasible at these higher bands through chip-scale arrays [11]. In fact, [12] shows 0% outage up to 200 m cell radius at 38 GHz for both a low (18 m) and high (36 m) transmitter (Tx) site. It should be noted that while diffraction is less important at these higher bands relative to the frequencies below 6 GHz, reflections are still present (but may behave differently than the lower frequencies) and also diffuse scattering is enhanced.

The new PL models for these higher frequencies should be consistent with the PL seen at the lower frequencies as well (e.g., from 3GPP [10]). However, it is clear that given the need to develop these PL models quickly to meet the near-term 5G standardization needs (e.g., in 3GPP), there will not be a complete set of measurements across a wide range of environments, distances, and frequencies. Hence the models will, out of necessity at least in the short-term, need to be applied in scenarios (e.g., frequencies, distances, or environments) where the model was not developed. Therefore it is important to understand how well different PL models predict the PL behavior outside of the scenarios where the measurements were taken.

In this paper we investigate how two different PL modeling methods behave when predicting the PL behavior to data outside of the data set used to compute the parameters of the PL models. In this paper the term "measurement set" will refer to the set of data used to compute the parameters of the PL model and the term "prediction set" will refer to a different set of data where the performance of the PL models will be compared. The two PL models considered in this paper are the close-in (CI) free space reference distance PL model [5][9] and the Alpha-Beta-Gamma (ABG) PL model [6][7]. The reason these two models are being compared is that they differ in their respective PL modeling methodologies. More specifically, one model, the CI PL model, is anchored to a physical reference distance, and the other model, the ABG PL model, has no anchor and is only a best fit curve match over the available data set. Hence it is valuable to compare the behavior of these two models outside of the measurement set (i.e., on the prediction set) to see if one or the other modeling methodology provides a more robust PL modeling both in terms of RMS error (aka shadow fading (SF) standard deviation) and in the stability of the parameters which make up the models. Note a more thorough discussion of the CI and ABG models is given in [8].

## II. OVERVIEW OF THE AALBORG MEASUREMENT DATA AND THE RAY TRACING STUDIES

In this paper the two PL models will be compared using both real-world measurements and results from a ray-tracing simulation. The measurement campaign was carried out at Aalborg, Denmark to investigate the propagation characteristics of the UMa environment at the super high frequency band [13]. The environment represented a typical European medium city's residential district, in which the building height and street width were relatively homogeneous and measured at 17 and 20 meter, respectively. There were 6 transmitter (TX) locations, and the TX heights were 20, 25 or 54 meters. A narrowband continuous wave (CW) signal was transmitted at the frequencies of interest, i.e. 10, 18 and 28 GHz, and another CW signal at 2 GHz was always transmitted in parallel and served as a reference. The receiver (RX) was mounted on a van, driving at a speed of 20 km/h within the experimental area. The driving routes were chosen so that they were confined within the 3 dB beamwidth of the TX antennas. The received signal strength and GPS location were recorded at a rate of 20 samples/s using the R&S TSMW Universal Radio Network Analyzer for the calculation of PL and TX-RX separation. The data points were visually classified into LOS and NLOS condition based on Google Maps. Interested readers can refer to [13] for further information on the setup and how the receiver antenna patterns were compensated for.

The ray-tracing simulation was performed using the WinProp v.13 ray-tracing simulator [14] with the 3D Standard Ray Tracing model (Fresnel coefficients for reflection and the uniform theory of diffraction (UTD) model for diffraction based on electrical parameters of materials, plus diffuse scattering was enabled). The simulation was carried out using two UMa environment models based on the Madrid-grid layout described in METIS project [15] which consisted of open squares and street canyons. In the first model, the original heights of buildings were used with three different TX locations and antenna heights of 51 m, 54 m and 46.5 m, respectively (i.e., these antennas were located on rooftops). In the second model, the buildings height was reduced to about 57% of the height in comparison to the original layout. The same three TX locations were used as in the first model but antenna height was now 29 m, 31 m and 27 m, respectively. The antenna height of the RX points were 1.5 m and isotropic antennas were used at both the TX and RX. The frequencies used in the simulation were the following: 2 GHz, 5.6 GHz, 10.25 GHz, 28.5 GHz, 39.3 GHz and 73.5 GHz. The walls and ground were modeled by electrical parameters for concrete for all frequencies according to ITU-R recommendation P.2040 [16]. The maximum numbers of reflections used in the simulations were 4 and maximum numbers of diffractions were 2 for frequencies below 10 GHz and 1 above 10 GHz. Only outdoor simulations were performed, the transmission from outdoor to indoor was disabled and 20 rays were calculated per RX point.

Only the non-line-of-sight (NLOS) data from both the measurements and ray tracing will be used in this study. Note that the ray tracing data includes points that would not be detectable in normal measurement campaigns due to the limited dynamic range of actual measurement equipment. Thus we only use ray tracing data where the PL minus free-space path loss (FSPL) at 1 m is less than 100 dB. Using this relative criterion makes the assumption that the sensitivity of the measurements improves with frequency (e.g., using more directional antennas at higher frequencies to improve gain) but has the benefit of having a relatively stable number of points for the different frequencies used in the ray-tracing study. Using an absolute criterion such as keeping data with PL below say 170 dB, besides making the assumption that the measurement equipment would be identical at all frequencies, results in less data being available at the higher frequencies. The concern was that the difference in the amount of ray tracing data at different frequencies could bias the results, and hence the relative criterion was adopted.

## III. Overview of the CI and ABG Path Loss models

The CI PL model is given as [8][9]:

$$\mathrm{PL}^{CI}(f,d)[dB] = \mathrm{FSPL}(f,1\,m) + 10n\log_{10}\left(\frac{d}{1\,m}\right) + X_\sigma^{CI}, \quad (1)$$

where $f$ is the frequency in Hz, $n$ is the PL exponent (PLE), $d$ is the distance in $m$, $X_\sigma^{CI}$ is the SF term, and the FSPL($f$, 1 $m$) is:

$$\mathrm{FSPL}(f,1\,m) = 20\log_{10}\left(\frac{4\pi f}{c}\right), \quad (2)$$

where $c$ is the speed of light.

The ABG PL model is given as:

$$\mathrm{PL}^{ABG}(f,d)[dB] = 10\alpha\log_{10}\left(\frac{d}{1\,m}\right) + \beta + 10\gamma\log_{10}\left(\frac{f}{1\,GHz}\right) + X_\sigma^{ABG}, \quad (3)$$

where $\alpha$ captures how the PL changes in distance, $\beta$ is an optimized offset value in dB, $\gamma$ captures how the PL changes in frequency, and $X_\sigma^{ABG}$ is the SF term.

In the CI PL model, only a single parameter, the PLE, needs to be determined and it can be found by minimizing the SF standard deviation over the data set [8]. What distinguishes the CI PL model is that there is an anchor point, the FSPL at 1 m, which captures the frequency-dependency of the PL and provides a physical basis for this model. Note that the 1 m anchor point makes physical sense even in NLOS conditions as even in NLOS the first 1 m around a typical transmitter is still obstruction-free and hence all rays still experience free-space propagation around the transmitter [5]. In the ABG PL model there are three parameters which need to be determined and they are chosen to minimize the SF standard deviation over the data set like the CI PL model [8]. However, since there are three parameters in the ABG PL model compared to only one in the CI PL model, the ABG model should always have a lower SF standard deviation than the CI PL model over the data set. However, as we will show in this paper, the lack of an anchor to physics will mean that when the ABG PL model is applied outside of the scenario (e.g., distance, frequency, or environment) where the data was taken, the SF standard deviation will tend to be higher than the CI PL model and the parameters of the ABG model are much more unstable than the PLE of the CI model.

The closed-form expressions given in the appendix of [8] which minimize the SF standard deviation are used to determine the PLE for the CI model and the three parameters of the ABG model. Over the entire set of data the two models had very similar SF standard deviations (8.95 dB for the CI PL model and 8.93 dB for the ABG model) indicating the extra parameterization of the ABG model does not significantly help improve the modeling accuracy in this particular measurement set. For reference, the PLE for the CI model was found to be $n$=2.67, and the ABG model parameters were: $\alpha$=2.62, $\beta$=34.90, and $\gamma$=1.90. So the two models, over the entire set of data, have very similar behavior across distance (similar $n$ and $\alpha$), but interestingly, the ABG model predicts that the PL (minus FSPL ($f$, 1 m)) would improve with frequency since $\gamma$<2.

## IV. Prediction in Distance

For all experiments we separated the data into two sets. The first set was the measurement set which was used to compute the parameters of the PL models and the second set was the prediction set where we computed the SF standard deviation (dB). For all experiments, we compute the SF standard deviation on the prediction set by computing the RMS error between the resulting models found using the parameters calculated from the measurements set to the measured path loss values on the prediction set. In this set of experiments we broke the total data set up into two portions based on distance. We kept the prediction set fixed in this investigation and varied the measurement set where the measurement set included distances which kept getting further away from the prediction set.

The first investigation is for the case that the prediction set is closer to the TX (base) than the measurement set. In this case the prediction set is all the data with distances less than or equal to $d_{\max}$ = 200 m, and the measurement set will vary as all distances greater than $d_{\max}+\delta_d$ ($\delta_d\geq0$). As the distance, $\delta_d$, between the two sets increases, it would be expected that the SF standard deviation of the two PL models would also increase. However, as can be seen in Fig. 1, the CI PL model had a very constant SF standard deviation for the prediction set regardless of how far away the measurement set would get. On the other hand, the ABG PL model's SF standard deviation on the prediction set increased substantially as the measurement set got farther away from the prediction set (i.e., as $\delta_d$ increased). Also, the stability of the PLE of the CI PL model is much higher than the parameters of the ABG model when varying the distance between the two sets as seen in Fig. 2. In particular, the $\alpha$ of the ABG model can vary quite a bit (1.53 to 2.73) which could have significant effects in system-level simulations as the level of interference seen greatly depends on the value of $\alpha$ (i.e., the distance-related parameter).

In the second investigation, the prediction set is for distances far from the TX (base) and the measurement set is close to the TX. In this case the prediction set is with distances greater than or equal to $d_{\min}$ = 900 m, and the measurement set will be variable as all distances less than $d_{\min}$-$\delta_d$ ($\delta_d\geq0$). The results for this case are shown in Fig. 3 and Fig. 4 for the SF standard deviation on the prediction set and the parameters of the PL models respectively, both as a function of $\delta_d$. In this case both PL models had a SF standard deviation that varied very little as the distance between the measurement set and prediction set increased, although the CI PL model did have a slightly lower SF standard deviation. Both models had parameters that were fairly stable with $\delta_d$, although the PLE of the CI PL model was slightly more stable than the parameters of the ABG PL model.

## V. Prediction in Frequency

In these experiments the prediction set will be the data for a given frequency and the measurement set all other frequencies. For example the prediction set could be all data at 2 GHz and the measurement set the data for all other frequencies. Note that when two frequencies are close to each other as is the case for 10 and 10.25 GHz and also for 28 and 28.5 GHz, they are both included in the 10 GHz and 28 GHz results, respectively.

Fig. 5 shows the SF standard deviation for the two PL models on the prediction and measurement sets for the frequency shown on the x axis (where that frequency is for the data in the prediction set). Both models seem to predict in frequency well except when predicting down to 2 GHz from the other frequencies where the CI PL model does a significantly better job at predicting the PL performance. Fig. 6 shows how the parameters of the PL models change when different frequencies are used in the prediction set. As can be seen, the CI PL model has a much more stable parameterization than the ABG model (e.g., $n$ of the CI PL model varies only from 2.65 to 2.68, whereas $\alpha$ of the ABG model varies from 2.43 to 2.67). Also the $\gamma$ value of the ABG model can be as low as 1.51 (i.e., when excluding 2 GHz data from the measurement set) indicating a strong decrease in the PL (minus FSPL(f, 1 m)) as the frequency increases which was not supported by the data (in general it is very consistent across frequency indicating a value of $\gamma$ around 2.0).

## VI. Prediction Across Environments

In these experiments we divided the total data set up into two portions based on the environment. The first set is the measurement set which was used to compute parameters of the PL models based on the one type of environment (Aalborg or Madrid-grid) and the second set is the prediction set used to compute the SF standard deviation. For example the measurement set could be all data from the Aalborg measurements and prediction set the data from Madrid-grid simulation. The calculation in the prediction set was performed for every frequency used in the measurement or simulation and two sets of antenna heights: *lowTX* (20m/25m in Aalborg and 29m/31m/27m in Madrid-grid) and *highTX* (54m in Aalborg and 51m/54m/46.5m in Madrid-grid).

In the first investigation, the data from the Madrid-grid is used for the measurement set and the prediction set is for data from the Aalborg environment. The SF standard deviation for the measurement set is 9.11 dB for the CI PL model and 8.87 dB for the ABG PL model indicating that the ABG PL model has better accuracy (but only by 0.24 dB). The SF standard deviation of the PL models for the prediction set is shown on Fig. 7. In this case the SF standard deviation of the CI PL model is lower in more cases than the ABG PL model (i.e., the ABG PL model is clearly better only for *Aalborg_2GHz_lowTX*). Note that the better prediction behavior of the CI PL model is especially visible in the case of campaigns with high TX antennas.

In the second investigation, the data from Aalborg is used for the measurement set and the prediction set is for data from Madrid-grid environment. The SF standard deviation for the measurement set in this case is 8.72 dB for the CI PL model and 8.58 dB for the ABG PL model indicating again that the ABG PL model has a slightly better accuracy (but only by 0.14 dB). The SF standard deviation of the PL models for the prediction set is shown on Fig. 8. In this case the SF standard deviation is better for the CI PL model for all cases. An interesting observation is that the CI PL model is much better in predicting PL values for the higher frequency bands which were not used in the measurement set (39.3 and 73.5 GHz). In this case the CI PL model is better in terms of SF standard deviation relative to the ABG PL model from 0.84 dB to 1.94 dB.

The presented results show that the CI PL model has a better prediction ability in most of the cases in term of SF standard deviation on the different environment which was not used for determining the PL model parameter. This better prediction ability was despite the ABG PL model having a slightly better SF standard deviation for the measurement set. This advantage is especially useful for near-term 5G standardization needs where a complete set of measurements across environments could be limited.

## VII. Conclusion

This paper presented a comparison of two path loss models in a UMa environment using measured data from 2 to 28 GHz and ray tracing data from 2 to 73.5 GHz. One path loss model had a physically-significant anchor point (CI model) and the other was a "floating" model (ABG model) which just does a curve fitting to the available data. It was shown that having the anchor point tied to physics improves both the stability of the model and the SF standard deviation seen when using the model to predict path loss at different distances, frequencies, and environments relative to the set of data where the parameters of the path loss models were originally determined. Thus, for unexpected scenarios or for situations where a path loss model must be used outside of the range of measurements used to create the original model, this paper shows the CI model is more robust and reliable as compared to the ABG model.

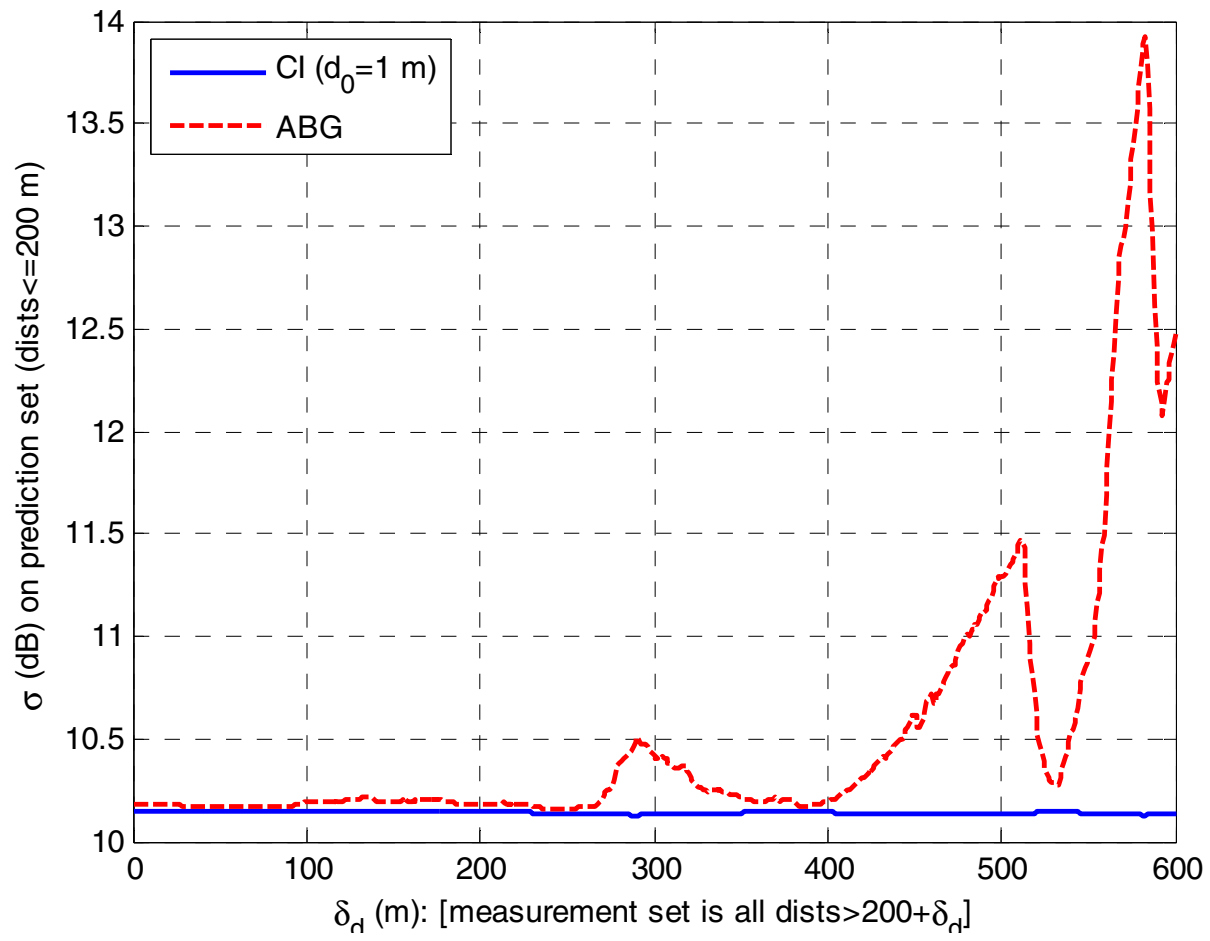

Fig. 1. SF standard deviation of the PL models for prediction in distance when the prediction set is close to the TX.

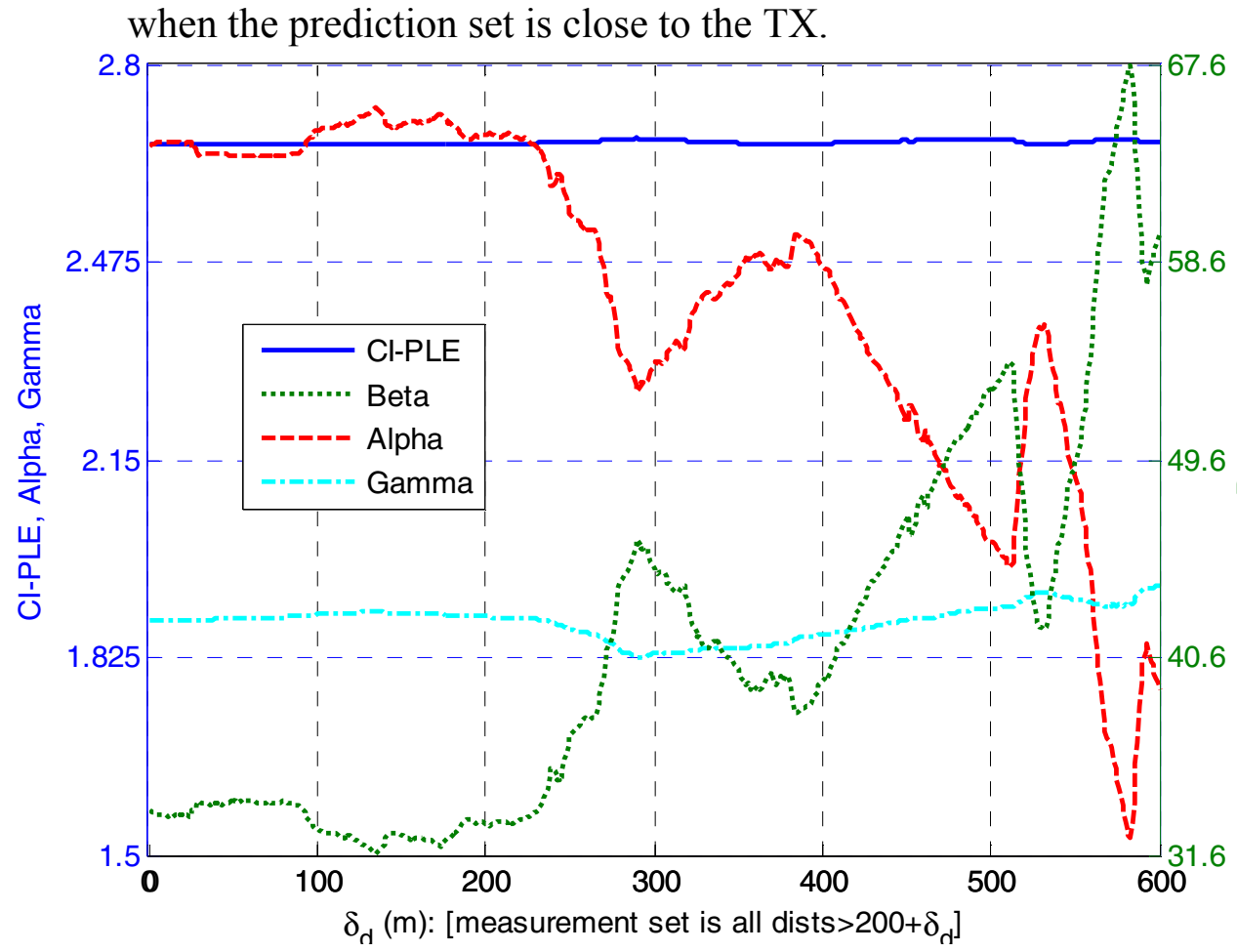

Fig. 2. Parameters of the PL models for prediction in distance when the prediction set is close to the TX. Note that the scale for beta is to the right.

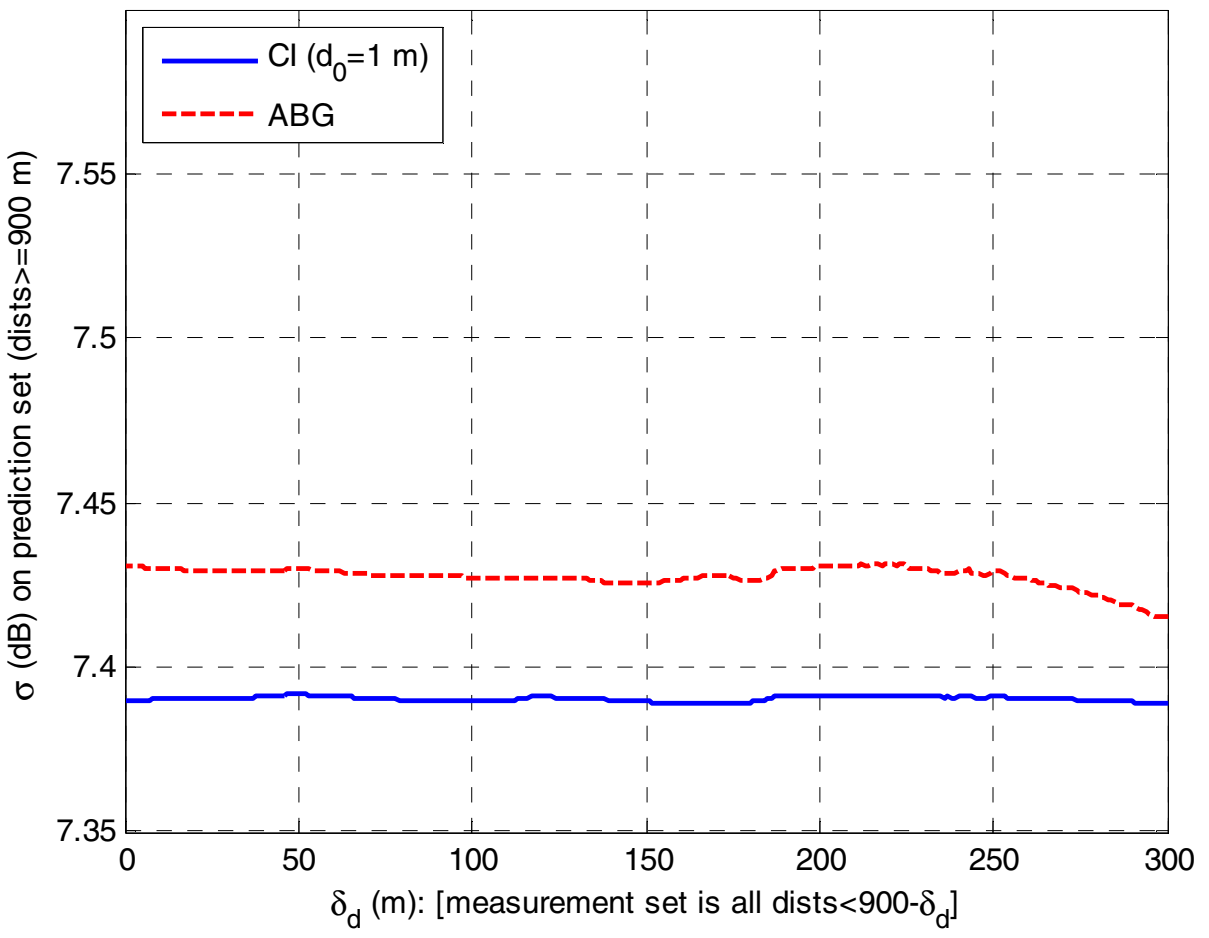

Fig. 3. SF standard deviation of the PL models for prediction in distance when the measurement set is close to the TX.

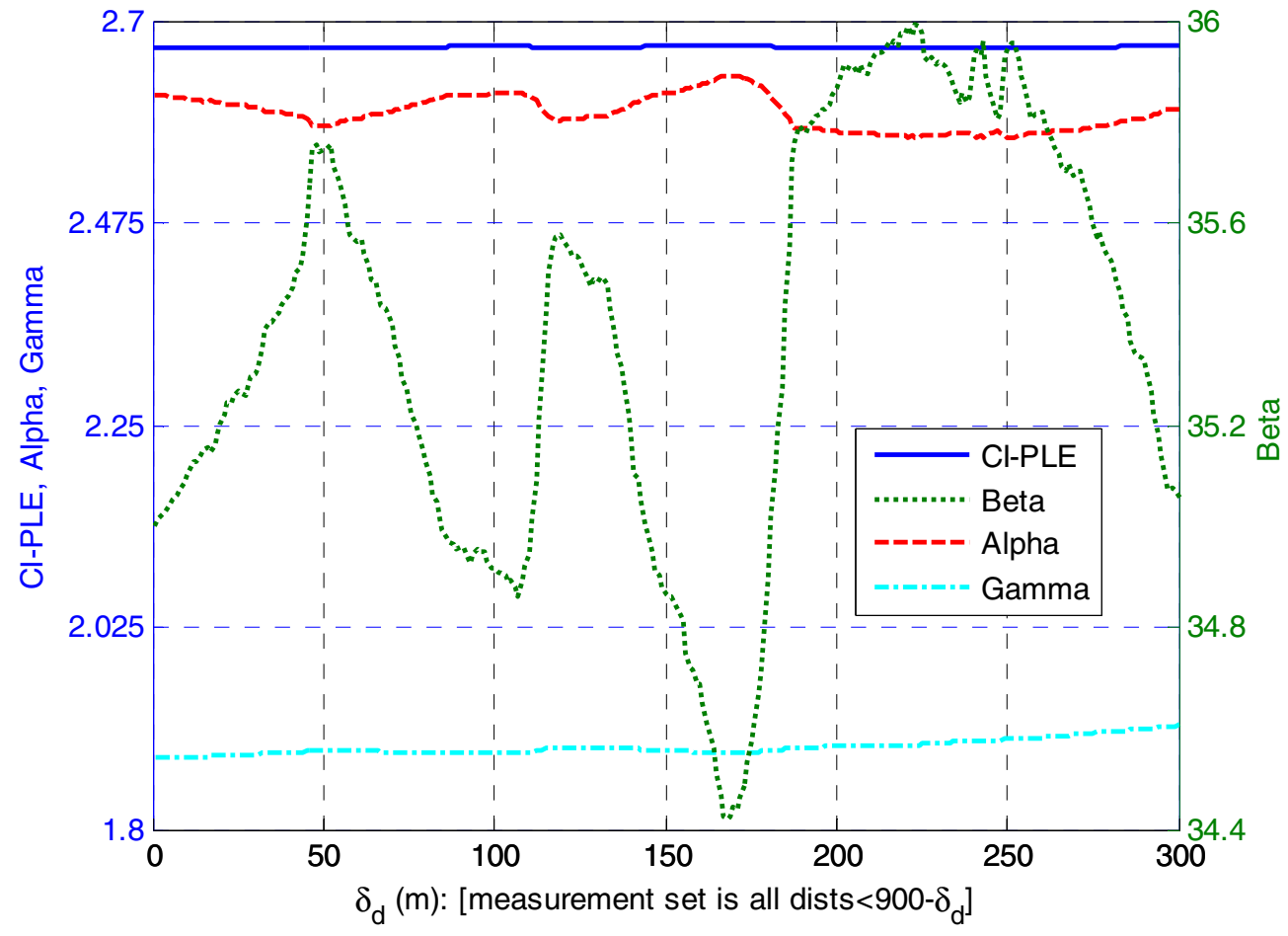

Fig. 4. Parameters of the PL models for prediction in distance when the measurement set is close to the TX. Note that the scale for beta is to the right.

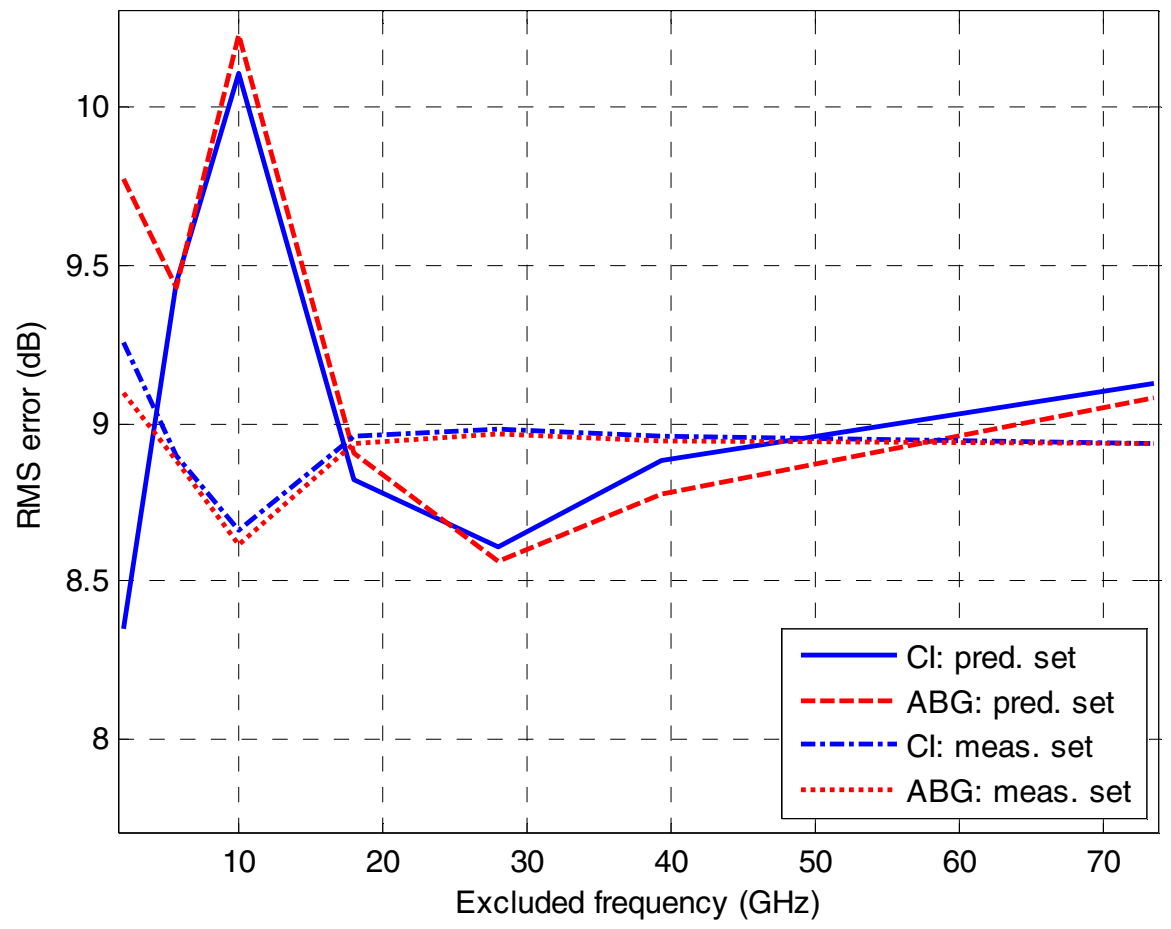

Fig. 5. SF standard deviation for the PL models for prediction in frequency at the frequency shown. The measurement set was for all frequencies except the excluded one shown on the x axis which is the prediction set.

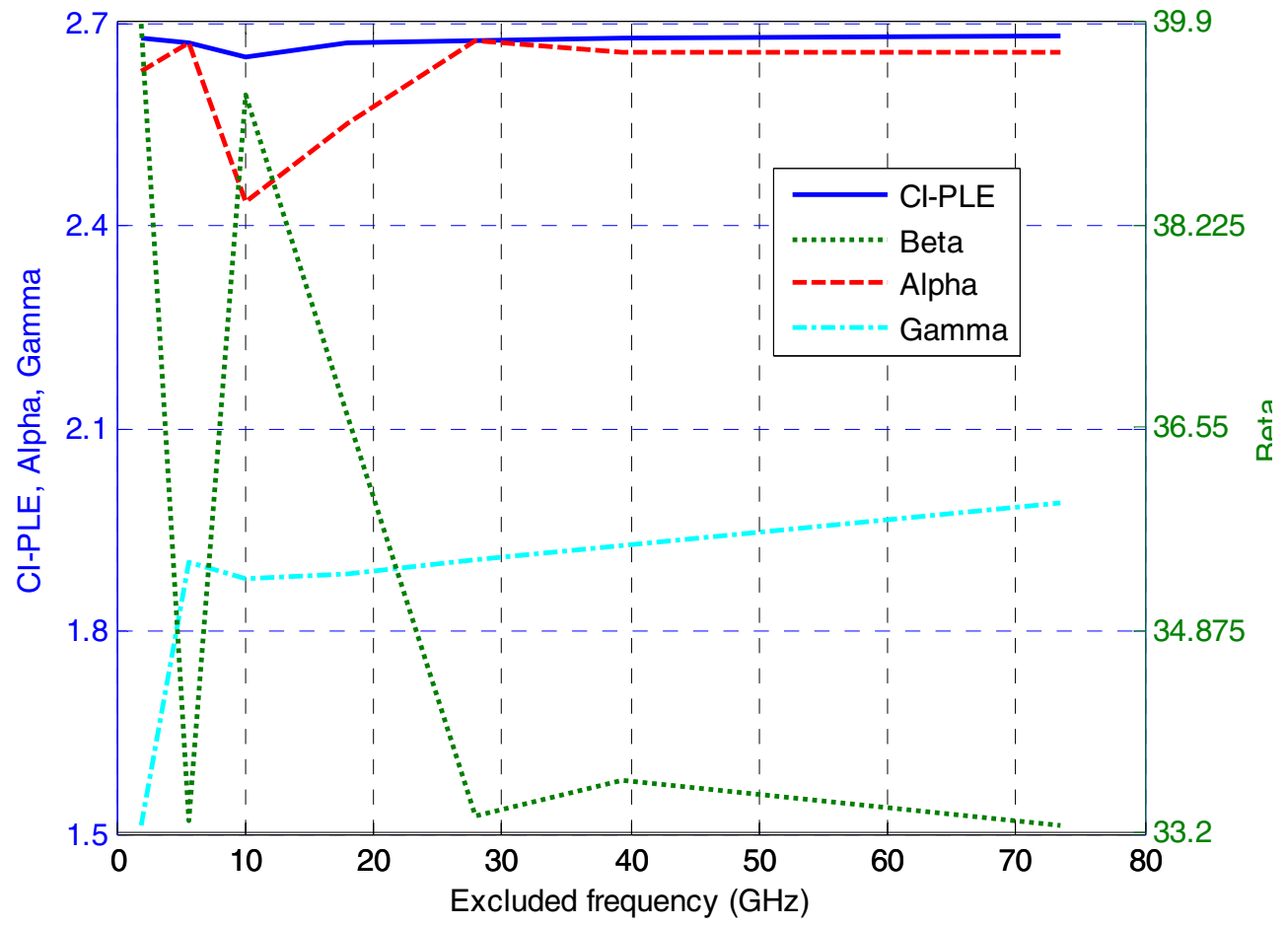

Fig. 6. Parameters of the PL models for prediction in frequency at the frequency shown. Note that the scale for beta is to the right.

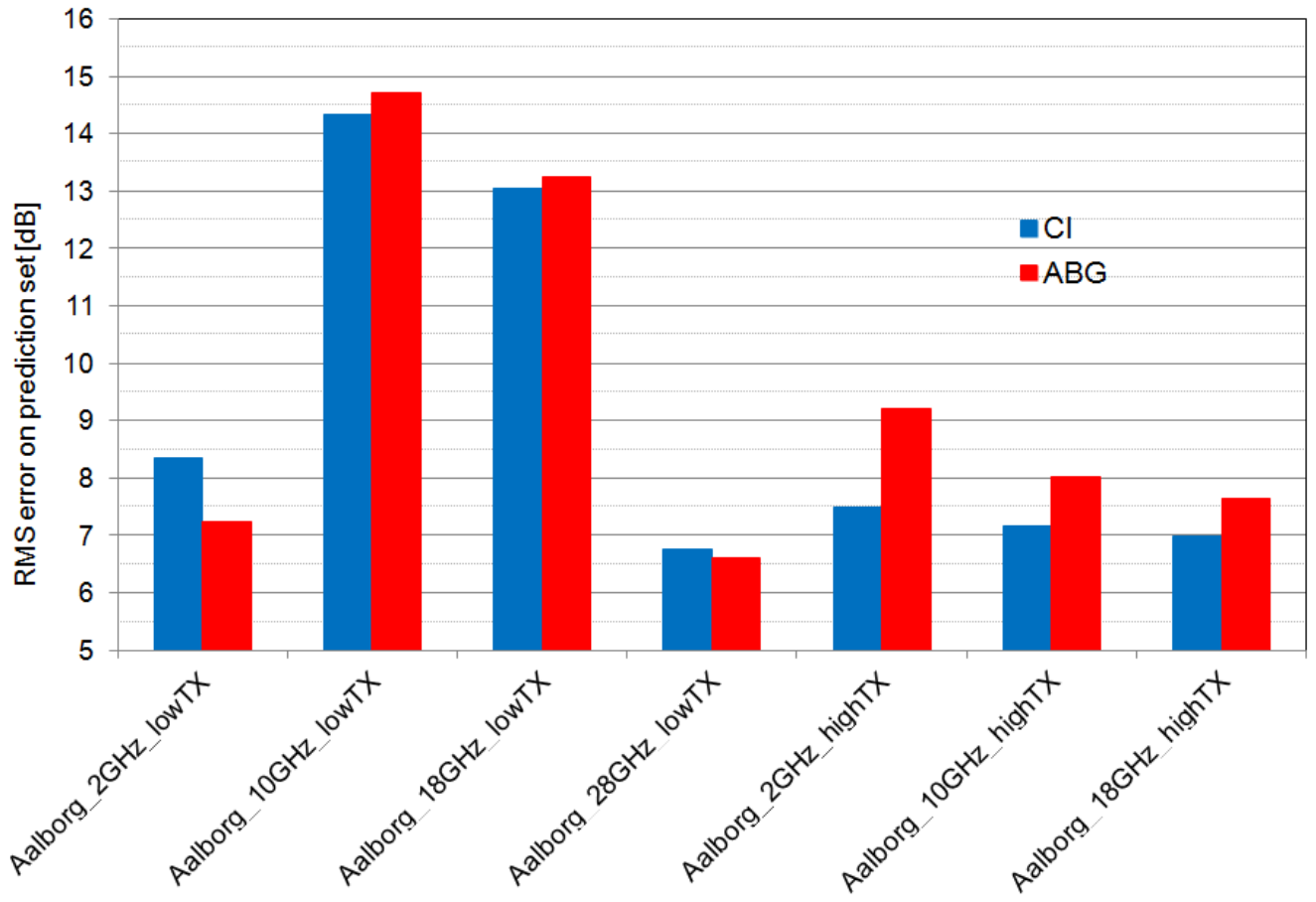

Fig. 7. SF standard deviation of the PL models when the prediction set is for data from the Aalborg environment and the measurement set is for data from the Madrid-grid.

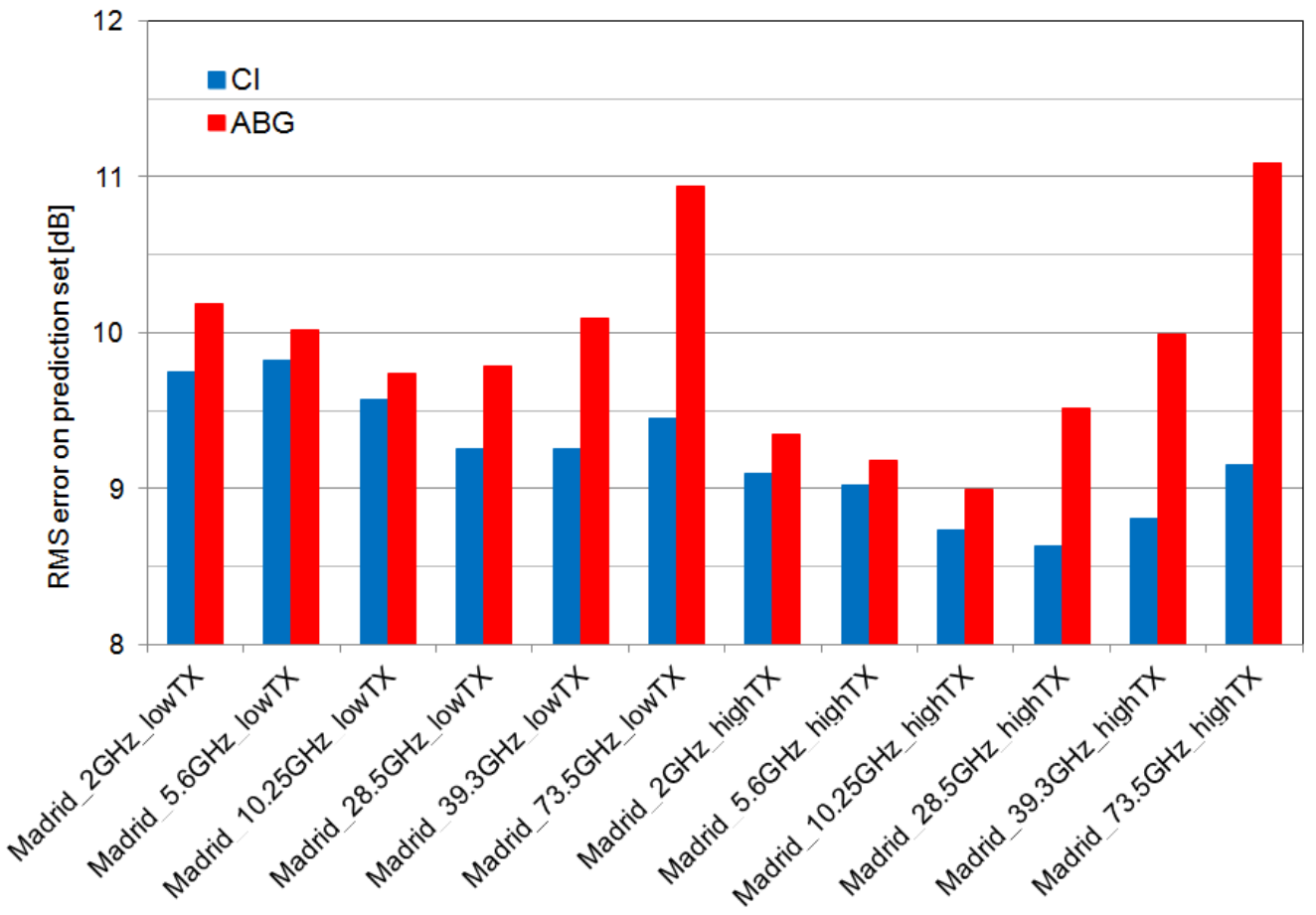

Fig. 8. SF standard deviation of the PL models when the prediction set is for data from Madrid-grid environment and the measurement set are data from Aalborg.